\documentstyle[conf_iap,epsf]{article}
\def\plottwo#1#2{\centering \leavevmode
\epsfxsize=.4\columnwidth \epsfbox{#1} \hfil
\epsfxsize=.4\columnwidth \epsfbox{#2}}
\def\hmpc{{\rm\, h^{-1}Mpc}}\def\br{{\bf r}}
\def\bk{{\bf k}}   
\def\hk{{\hat k}}    
\def\spose#1{\hbox to 0pt{#1\hss}}
\def\lta{\mathrel{\spose{\lower 3pt\hbox{$\mathchar"218$}}
     \raise 2.0pt\hbox{$\mathchar"13C$}}}
\def\gta{\mathrel{\spose{\lower 3pt\hbox{$\mathchar"218$}}
     \raise 2.0pt\hbox{$\mathchar"13E$}}}
\def\etal{{\it et al. }}

\begin{document}
\heading{Cosmic Web: Origin and Observables}
\par\medskip\noindent

\author{Dmitry Pogosyan$^{1}$, J. Richard Bond$^{1}$, Lev Kofman$^{1}$ and James Wadsley$^{1,2}$}
\address{
Canadian Institute for Theoretical Astrophysics, University of Toronto,\\
\mbox{\hspace{3mm}}60 St. George St., Toronto, ON M5S 3H8, Canada}
\address{Astronomy Department, Univ. of Washington,  Box 351580, Seattle WA 98195-1580}
\begin{abstract}
Simple analytic arguments are used to understand the predominantly
filamentary web in the large-scale distribution of galaxies.
Numerical simulations of superclusters are performed to assess the
feasibility of directly mapping the intracluster webbing, in
particular with weak gravitational lensing using Wide Field Cameras.
\end{abstract}
\section{Patterns in random density fields}
In \cite{bkp96} we have shown (see also \cite{bkpwhere} in this
volume) that the large-scale structure in N-body simulations, which is
believed to describe adequately the galaxy distribution in redshift
catalogues, can be explained directly by the geometry of the initial
density field, amplified by gravitational instability. While rich
clusters of galaxies form from high density enhancements of scales
$R_f\sim 8\hmpc$, intracluster filamentary webbing reflects initial
connections in the random density field bridging these protocluster
regions at lower density thresholds.

The quantitative study of the properties of 3D random fields in a
cosmological context goes back to the famous formula of Doroshkevich
\cite{d} for the joint distribution function
$P(\lambda_1,\lambda_2,\lambda_3) $ of the ordered eigenvalues
$\lambda_1\le\lambda_2\le\lambda_3$ of the initial deformation (shear)
tensor $e_{ij}$. The latter is related to the density field, given in
Fourier space by (random) amplitudes $\delta(\bk)$, by
\begin{equation}
e_{ij} (\br) =
-\int {d^3{\bk} \over (2 \pi)^3} \; \hk_i \hk_j \delta(\bk) 
W(kR_f) \, e^{i\bk\cdot \br} \, .
\end{equation}
The filter $W(kR_f)$ separates the smooth large-scale field which
evolves in the mildly nonlinear single-stream regime from the highly
nonlinear short-scale chaotic flow.

Usually, the dynamical role of the shear in a local gravitational
collapse is being focused on.  Here we stress another aspect of the
shear tensor: as a shape parameter describing the {\it geometry} of
the initial density field.  The mean overdensity profile around a
field point ${\bf r_0}$ with known values of the shear,
$\langle\delta({\bf r}) | \lambda_1({\bf r_0}),\lambda_2({\bf
r_0}),\lambda_3({\bf r_0})\rangle$, where $\delta =\rho/{\bar
\rho}-1$, as illustrated in Fig.~1, shows that the shape depends on
the sign of the shear eigenvalues (see \cite{p1} for details of these
and subsequent analytical calculations, also \cite{bbks}).  In effect,
the local shear signature defines the curvature of the density
isocontours up to a distance of several $ R_f $.\footnote{The
information contained in the density curvature tensor $\delta_{,ij}$
itself is much more local and less representative of the density
behaviour at large distances from the constraint point.}
\begin{figure}
\centerline{\epsfxsize=4.5in\epsfbox{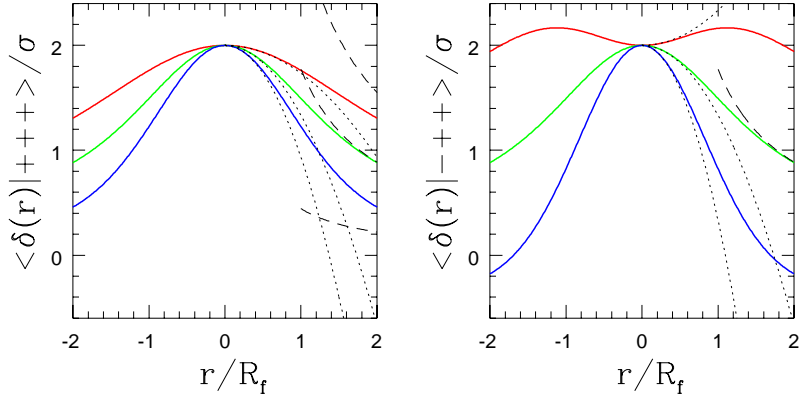}}
\caption{\label{fig:shape3} Constrained primordial density field
$\langle\delta({\bf r}) | \lambda_1,\lambda_2,\lambda_3\rangle$ as
a function of distance ${\bf r}$ in units of the filter scale $R_f$, in
the three eigendirections (1-top, 3-bottom).  The left plot corresponds to
the shear with all positive (+++) eigenvalues. The right plot
represents the case of "filamentary" behaviour of the density in the
neighbourhood of the ({--}\,++) sheared point. Dotted and dashed curves
show the analytic short and long distance asymptotics \cite{p1}.}
\end{figure} 
The increase in the density along one axis while falling off along the
remaining two is symptomatic of filamentary bridges that connect the
higher density regions where the shape of the density profile is more
spherical. Fig.~2 shows that local properties of the shear tensor
determine whether the region is part of the filamentary or
protocluster structure.
\begin{figure}
\centerline{\epsfxsize=4.5in\epsfbox{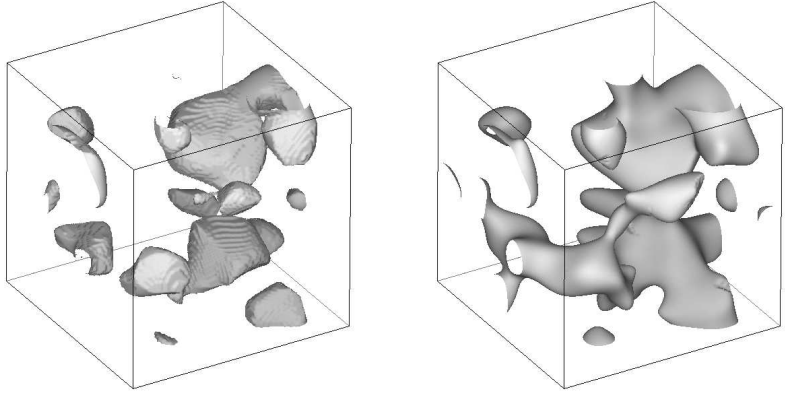}}
\caption{\label{fig:l3l2} Example of the cosmological
initial density field. Left panel: the part of space with
overdensity $\delta \ge 1\sigma$ and $\lambda_3 \ge \lambda_2 \ge \lambda_1 > 0$.
Note the roughly spherical shapes of the regions drawn. Right panel:
the overdense regions with $\delta \ge 1\sigma$ and
$\lambda_3 \ge \lambda_2 > 0$ but arbitrary $\lambda_1$. Inclusion of the
points with $\lambda_1 < 0$ creates filamentary connectors between round
$\lambda_1 > 0$ regions.}
\end{figure} 

This correspondence between the global properties of the density field
and local characteristics of the shear tensor enables analytical
estimation of the gross properties of the Cosmic Web.  For instance,
the probability of the eigenvalue signature given the density
threshold $\delta$, $P({\rm sign}|\delta)$, shows which structures
dominates depends on the isodensity threshold chosen. Fig.~3 shows that
at overdensities above a critical $\delta$, 1.56$\sigma$ for Gaussian
fields, one encounters predominantly spherical-like mass
concentrations, while at the lower density contrast,
$0<\delta<1.56\sigma$, most of the initial density enhancements are in
elongated filamentary bridges.  Planar configurations ({--}\,{--}\,+)
are less likely for any positive overdensities $\delta > 0$.
\begin{figure}
\plottwo{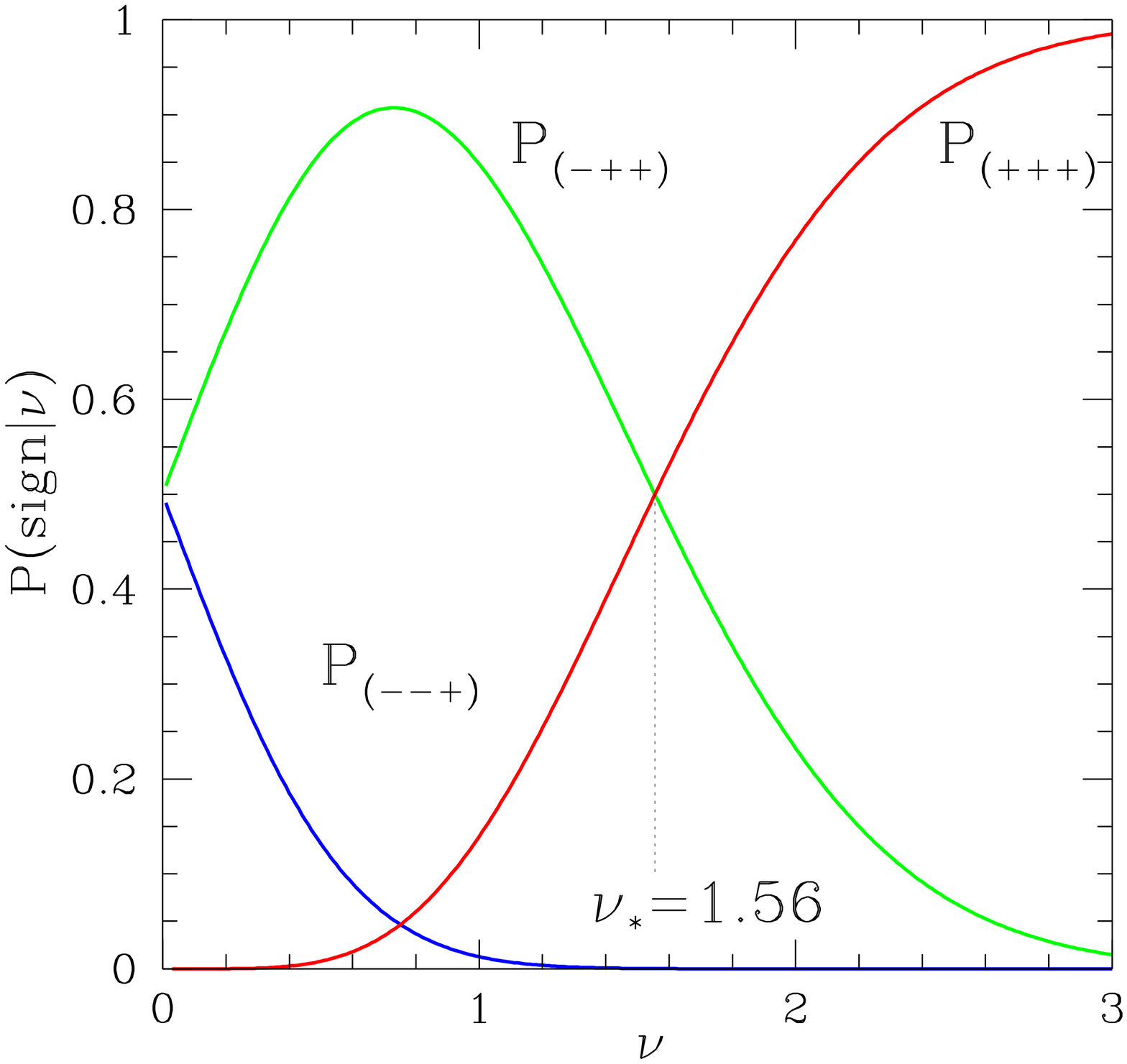}{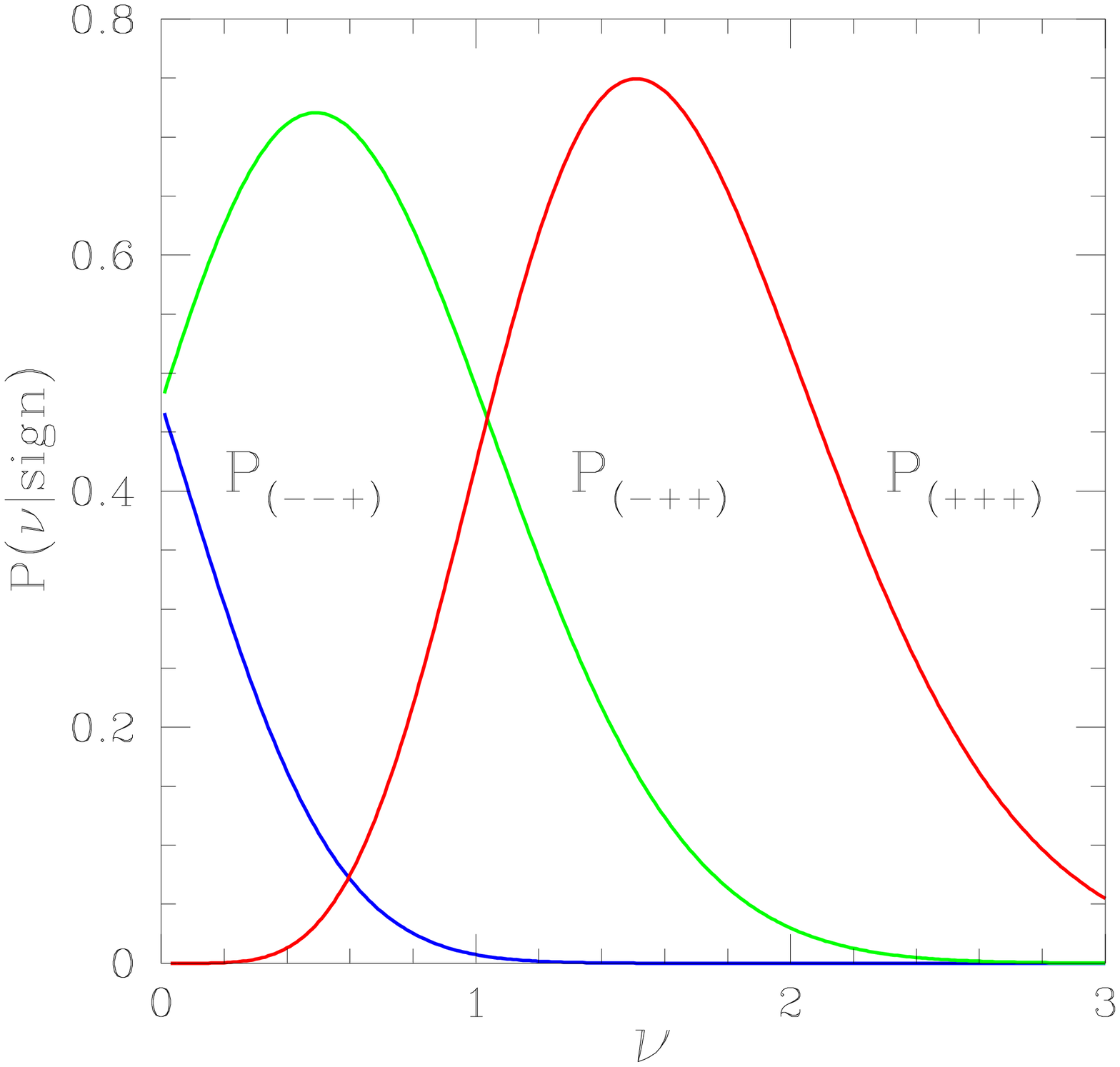}
\caption{\label{fig:filprob} Left panel: Probablity of the eigenvalue
signature given the overdensity threshold $P({\rm sign}|\nu)$,
$\nu=\delta/\sigma$.  Right panel: Density distribution given the type
of shear tensor, $P(\nu|{\rm sign})$.  Mean densities for the given
shear type are $\bar \delta_d \approx 1.66\sigma,0.6\sigma,-0.6\sigma$
for $(+++)$,$(-++)$ and $(--+)$ cases labeled by $d=3,2,1$ with
dispersion $\sigma_d\approx 0.55 \sigma$ almost equal for all
configurations.}
\end{figure} 
The related quantity $P(\delta|{\rm sign})$ gives us the density
distribution within different types of structure. While the average
density of the filaments in the initial configuration is equal to
$0.6\sigma$ (Fig.~3), it is the 1.5--2$\,\sigma$ excursions which are
precursors of the rare prominent filaments.
In contrast, even rare planar, membrane-like,
configurations are expected only at lower overdensities of 0.5--1$\,\sigma$.

As a simple nonlinear model for evolution of clusters,
filaments and membranes, let us consider the evolution of a shell which is spherical ($d$=3) or
cylindrical ($d$=2) of radius $R(t)$, or planar ($d$=1) of width
$2R(t)$ which expands with the Hubble flow in the other $3-d$
directions  (c.f. Fillmore and Goldreich \cite{fg}).  The
shell is characterized by its Lagrangian tophat scale $R_f$ and the
linear overdensity within the shell $\delta_d$. Evolution of the shell
radius in different cosmologies with scale factor $a(t)$ is given by
\footnote{$\Omega_m$ and $\Omega_{\Lambda}$ are present-day 
density parameters of the matter and the cosmological constant. 
$D(t_i)$ is the growing mode
at the onset of integration, $\dot R(t_i) = 
\dot a(t_i) \left[1-D(t_i) \delta_d/d \right] R_f $}
\begin{equation}
\ddot R = H_0^2\left[ {(3-d) \over 2d}~ \Omega_m +  a^3~ \Omega_{\Lambda} 
 - {3 \Omega_m \over 2 d } (1+D(t_i)\delta_d)
\left({a R_f\over R}\right)^d \right] {R \over a^3} \ .
\label{cosmol1}
\end{equation}
As we have shown, for cosmological initial conditions typical linear
overdensities $\delta_d$ at the same scale $R_f$ are quite distinct
for structures of different geometry.  Solving Eq.(\ref{cosmol1}) with
$\delta_d$ suggested by Fig.~3 demonstrates the range of present time
densities one can expect for objects of different shape on the basis
of this elementary theory.  Results are presented in Fig.~4.
\begin{figure}
\centerline{\epsfxsize=4.0in\epsfbox{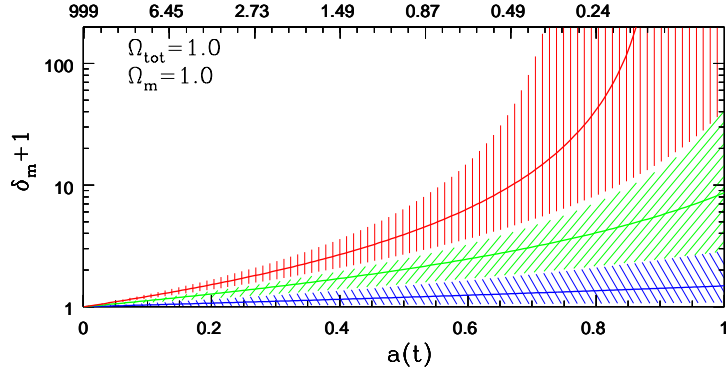}}
\caption{\label{fig:elementary} The nonlinear density evolution
$\delta(t)$ of idealized spheres (top curve), filaments (middle) and
membranes (bottom) of Lagrangian scale $R_f$ for which the smoothed
linear {\it rms} amplitude of density fluctuations $\sigma(R_f)$ is
$0.65$.  Solid curves show $\delta(t)$ for the structures with initial
(linear) density equal to $\bar\delta_d+2 \sigma_{d}$, $d=1,2,3$.  The
shaded bands delineate the range of $\delta(t)$ for initial densities
$[\bar \delta_d + \sigma_{d}, \bar \delta_d + 2.5 \sigma_{d}]$.
Values of $\bar\delta_d$ and $\sigma_{d}$ are given in Fig.~3.  }
\end{figure} 
For filaments we have $\rho \approx\,$3--12$\,\bar \rho$.  Although
filamentary structures have, generally, turned around, only rare
filaments
are approaching collapse by the present time.  Membranes, which are
still expanding, have $\rho \approx\,$1--3$\,\bar \rho$.
These values should be compared with the density of the collapsed
spherical halos at the moment of virialization, $\approx 180 \,\bar
\rho$. N-body simulations confirm these simple estimates.

\section{Observability of the Cosmic Web}
The filamentary pattern that joins galaxy clusters into a complex web
is manifest in 3D galaxy catalogues. However the question of biasing
between the galaxy distribution and the underlying mass density
remains open.  In recent work \cite{bkpw} we evaluated the feasibility
of mapping the intracluster structure using observational techniques
directly probing large-scale dark matter and gas distributions.  We
have performed gasdynamical simulations (using a SPH+TreeP$^3$M code
\cite{bwprep}) of several rich supercluster regions and considered the
weak gravitational lensing of high redshift galaxies by the low
redshift structure, the secondary variations in CMB temperature
induced by the Sunyaev-Zeldovich effect and the X-ray emission by hot
gas in clusters and along the filaments. The simulations and SZ maps
derived from them are described in our companion paper
\cite{bkpwhere}.

Since the lensing depends upon the projected density $\Sigma$, not
pressure as for the SZ effect or density-squared as for the X-rays,
weak lensing would be a better probe for the filaments if the
signal-to-noise is large enough.  There are two sources of noise, one
the intrinsic (random) ellipticities of the source galaxies, and the
other shear from ambient large scale structures along the
line-of-sight confusing the ellipticity pattern created by the
filamentary web.

Results of our simulations for standard CDM, open CDM and a CDM model
with a cosmological constant are presented in Fig.~5. They demonstrate
that cluster-cluster bridges at redshifts $z \sim 0.5$ are
sufficiently developed to produce lensing of background galaxies at
the level of several percent which can be detected in the images of
modern wide field cameras.  One such detection of the bridge between
two close clusters in the field of three at $z=0.4$ is described in
\cite{kl}.

\begin{figure}
\centerline{\epsfxsize=5.0in\epsfbox{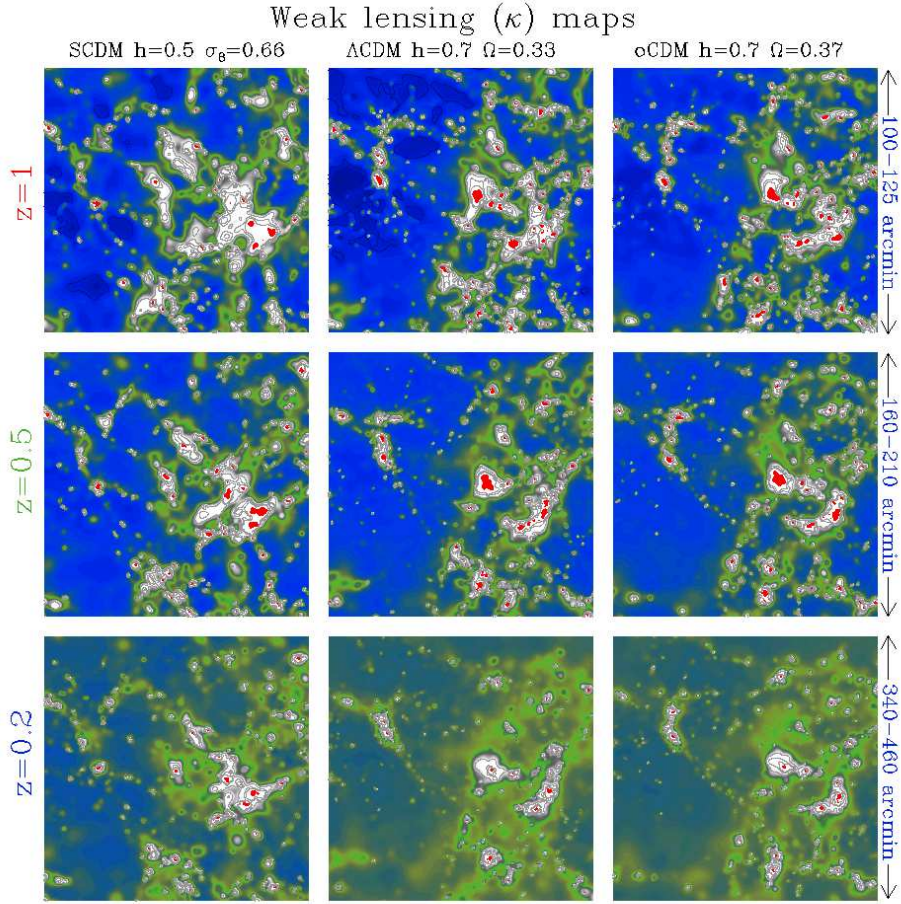}}
\caption{Colour maps of the convergence $\kappa=-2 \Sigma/\Sigma_{\rm crit}$
for our compact supercluster simulations at $z=0.2,0.5$ and $1$.
$\Sigma$ is the surface density (with the mean taken out),
and the critical density $\Sigma_{\rm crit}$ depends upon the comoving
distance to the lens and the source. We assumed the latter are at redshift $z=2$ for 
these plots. White areas correspond to $\kappa < -0.04$ with grey contours
marking the levels $-0.08,-0.12,-0.16$. Red dots  denote
high density regions $\kappa < -0.3$. The level of convergence in the green areas
is $\sim -0.02$. See Figs.~4,5 of {\protect\cite{bkpwhere}} for the
corresponding dark matter distributions and SZ maps.}
\end{figure} 

\acknowledgements{D.P wishes to thank the organizers of the XIV IAP Colloquium for
financial support.}


%
\vfill
\end{document}